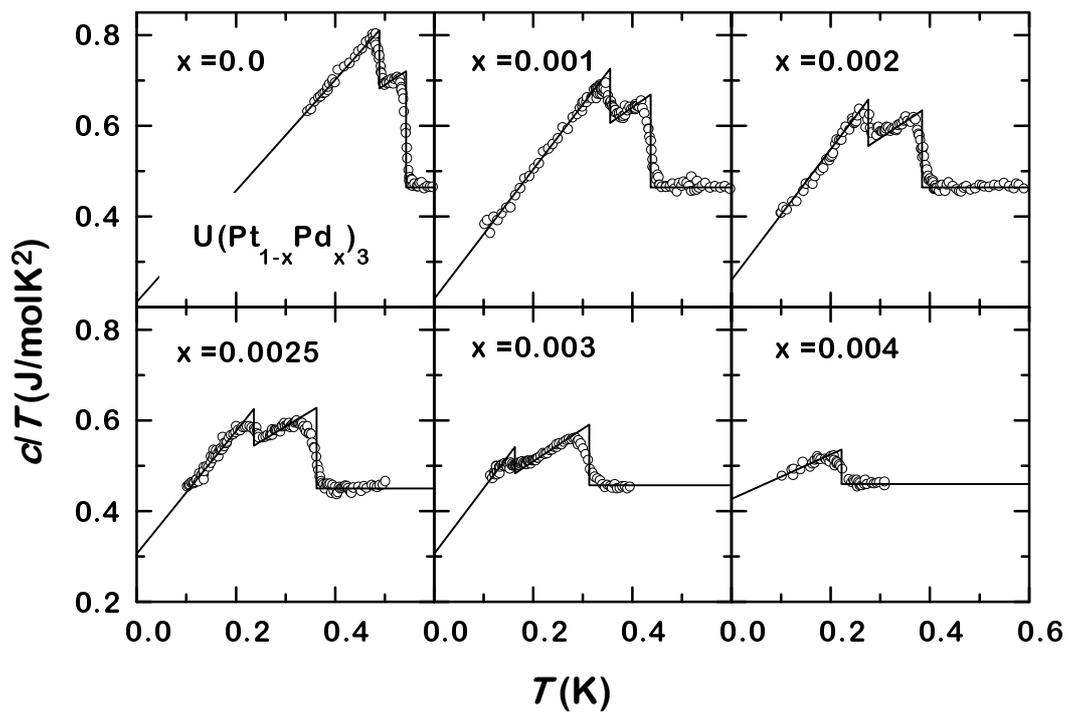

Fig. 1

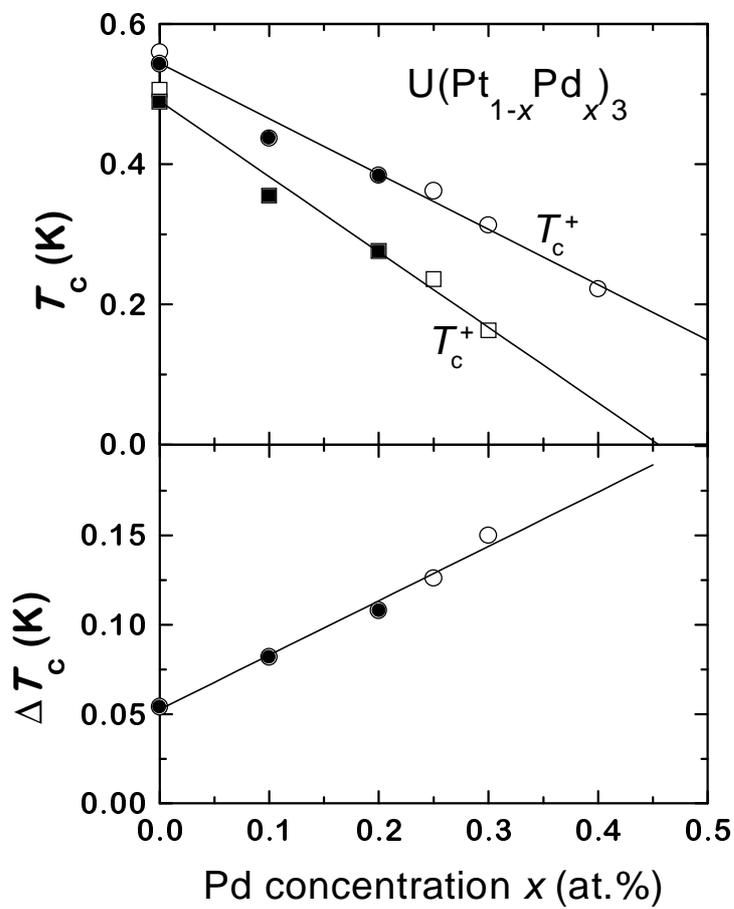

Fig. 2

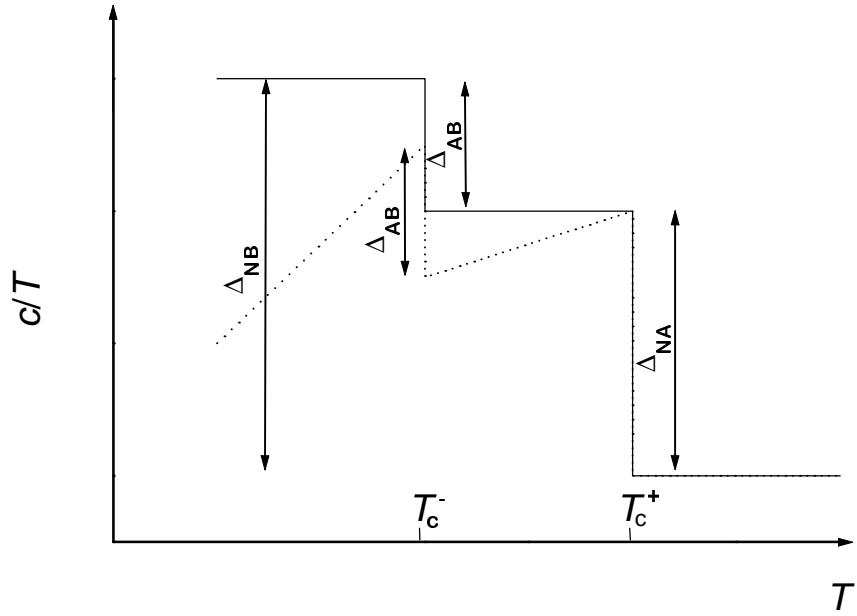

Fig. 3

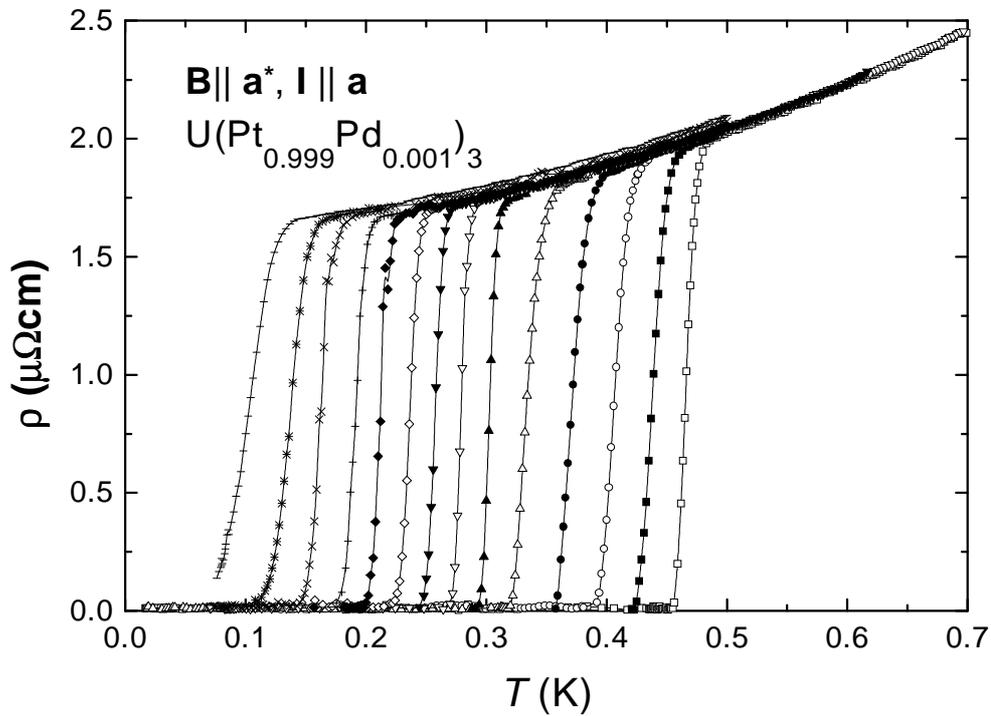

Fig. 4

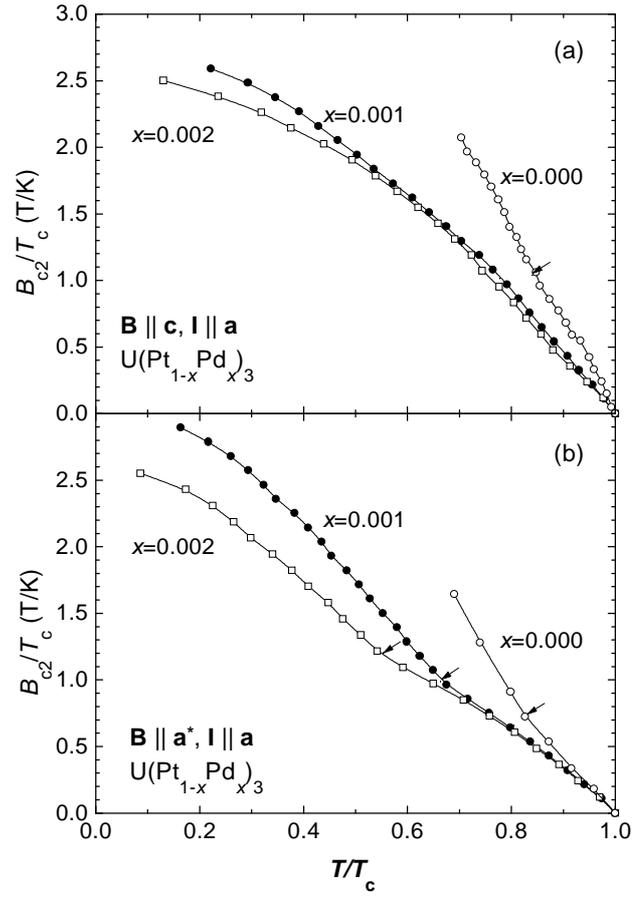

Fig. 5

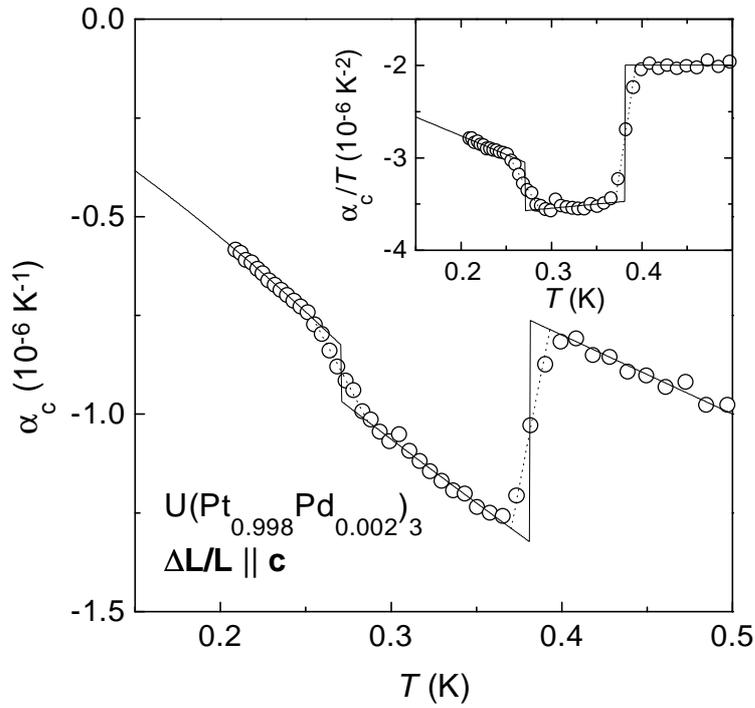

Fig. 6

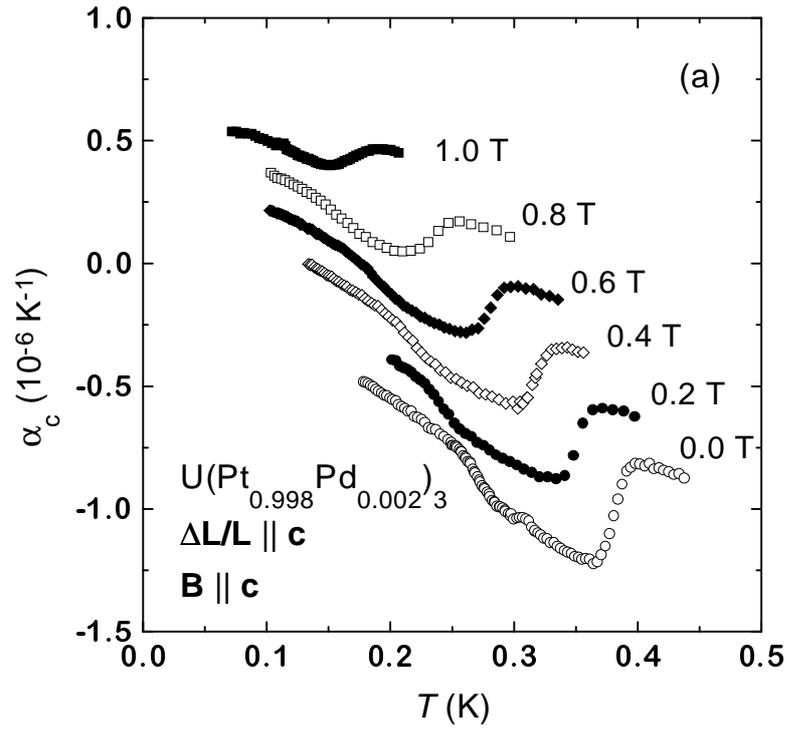

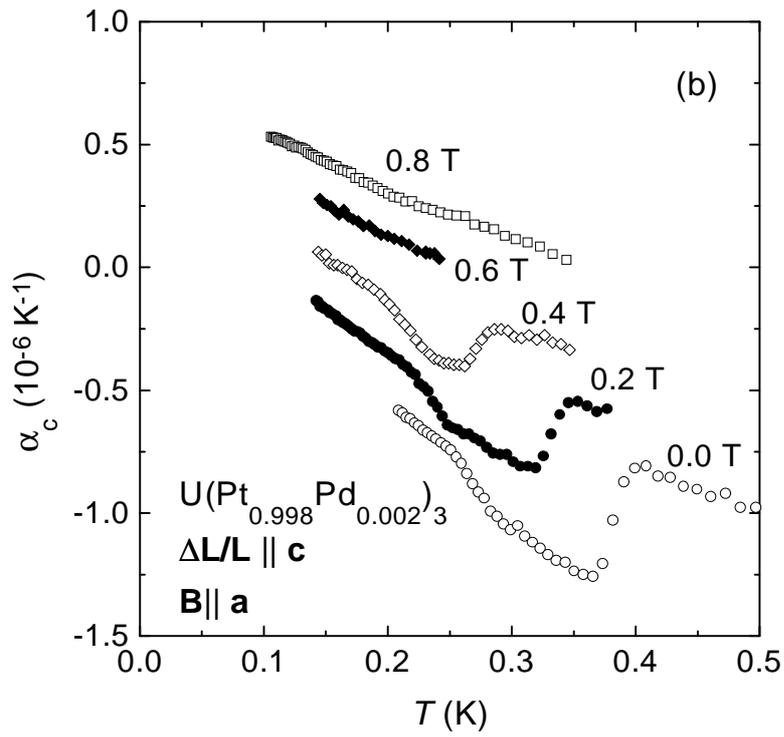

Fig. 7

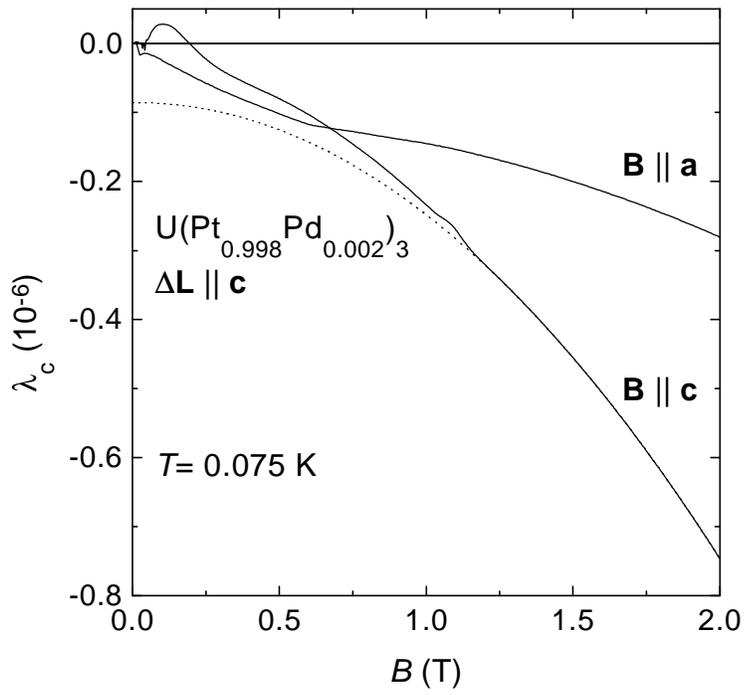

Fig. 8

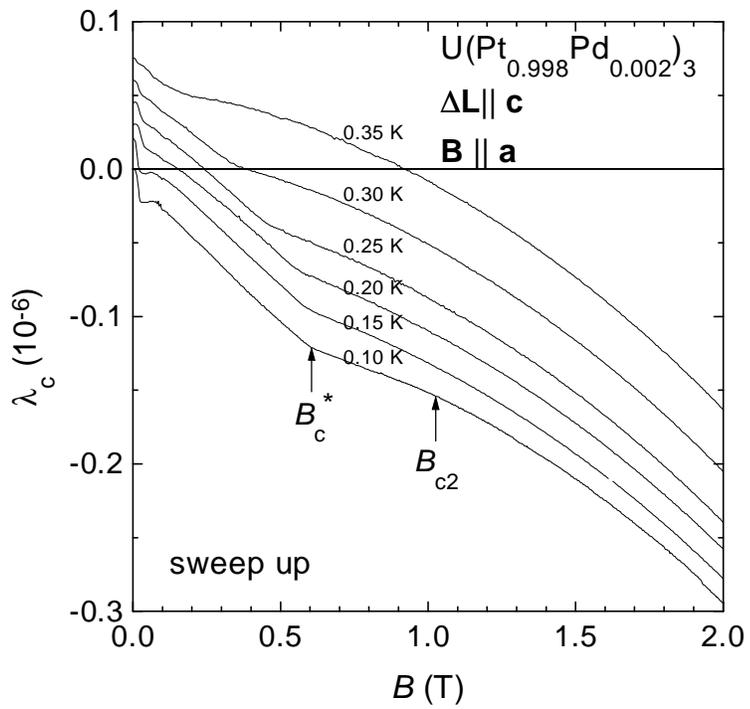

Fig. 9

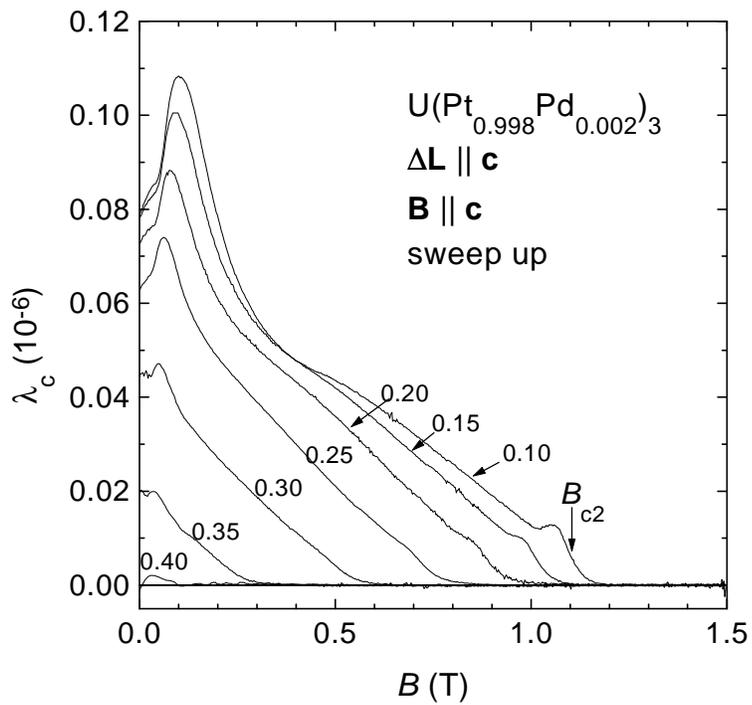

Fig. 10

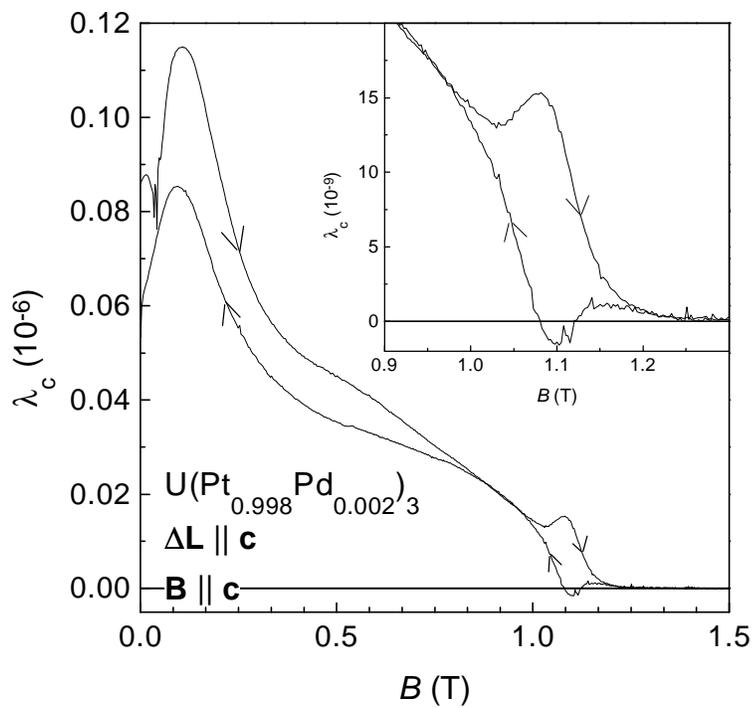

Fig. 11

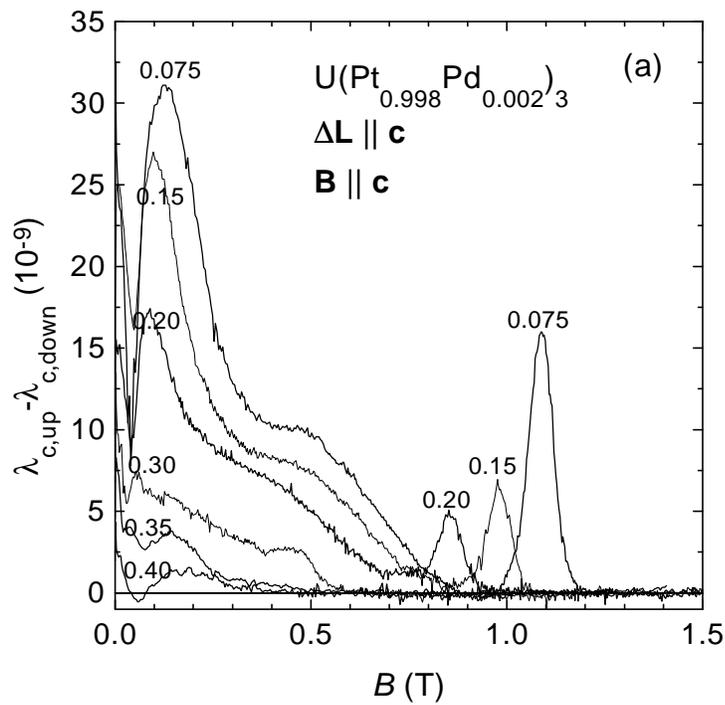

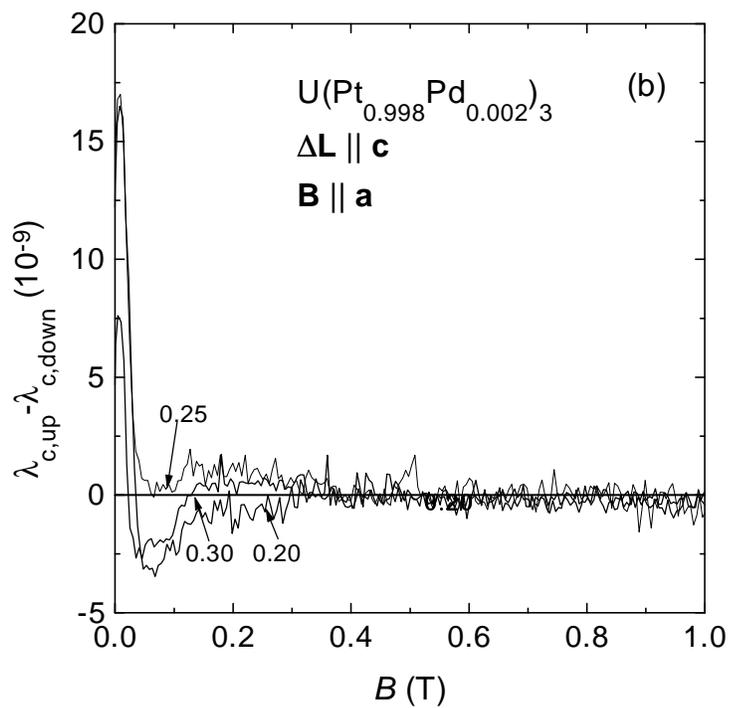

Fig. 12

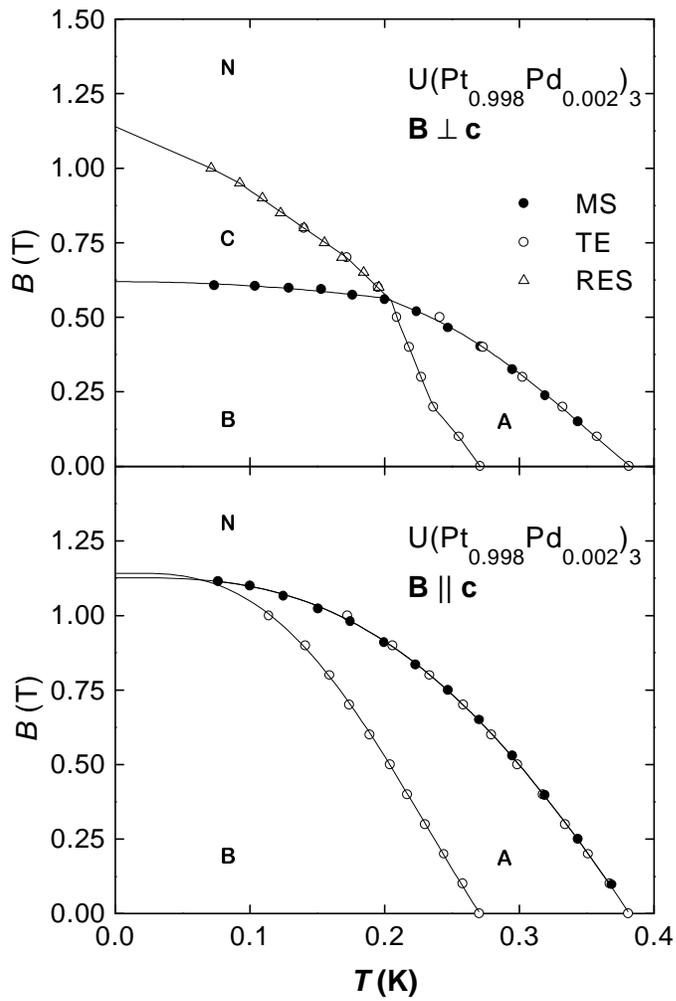

Fig. 13

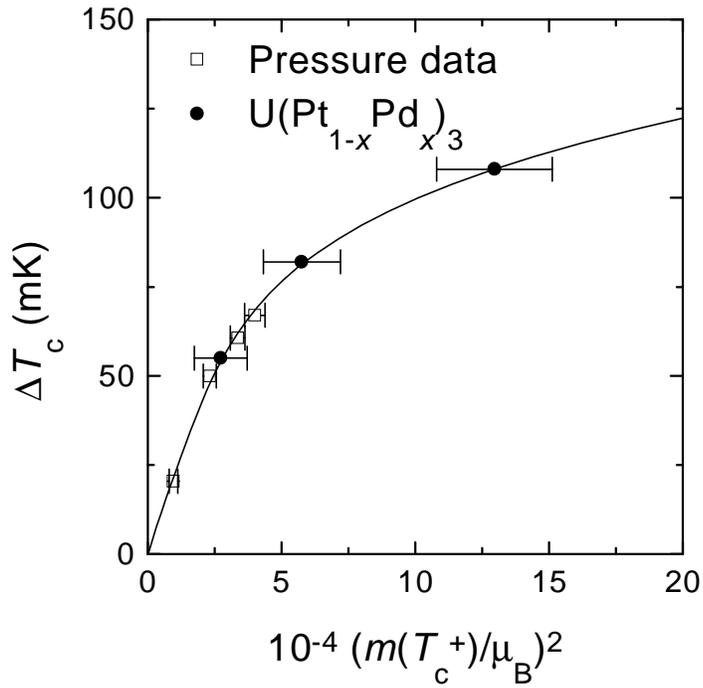

Fig. 14

# Superconductivity in heavy-fermion U(Pt,Pd)$_3$ and its interplay with magnetism

R.J. Keizer[1]*, A. de Visser[1], M.J. Graf[2], A.A. Menovsky[1], and J.J.M. Franse[1]

[1]*Van der Waals-Zeeman Institute, University of Amsterdam,
Valckenierstraat 65, 1018 XE Amsterdam, The Netherlands*
[2]*Department of Physics, Boston College, Chestnut Hill, MA 02467, USA*

## Abstract

The effect of Pd doping on the superconducting phase diagram of the unconventional superconductor UPt$_3$ has been measured by (magneto)resistance, specific heat, thermal expansion and magnetostriction. Experiments on single- and polycrystalline U(Pt$_{1-x}$Pd$_x$)$_3$ for $x \leq 0.006$ show that the superconducting transition temperatures of the A phase, $T_c^+$, and of the B phase, $T_c^-$ both decrease, while the splitting $\Delta T_c$ *increases* at a rate of $0.30 \pm 0.02$ K/at.%Pd. The B phase is suppressed first, near $x = 0.004$, while the A phase survives till $x \cong 0.007$. We find that $\Delta T_c(x)$ correlates with an increase of the weak magnetic moment $m(x)$ upon Pd doping. This provides further evidence for Ginzburg-Landau scenarios with magnetism as the symmetry breaking field, i.e. the 2D E-representation and the 1D odd parity model. Only for small splittings $\Delta T_c \propto m^2(T_c^+)$ ($\Delta T_c \leq 0.05$ K) as predicted. The results at larger splittings call for Ginzburg-Landau expansions beyond 4th order. The tetracritical point in the *B-T* plane persists till at least $x = 0.002$ for $\mathbf{B} \perp \mathbf{c}$, while it is rapidly suppressed for $\mathbf{B} \parallel \mathbf{c}$. Upon alloying the A and B phases gain stability at the expense of the C phase.

*Corresponding author:      R.J. Keizer
                            Van der Waals-Zeeman Institute
                            University of Amsterdam
                            Valckenierstraat 65
                            1018 XE Amsterdam
                            The Netherlands
                            Phone: 31-20-5255795
                            Fax:     31-20-5255788
                            E-mail: rjkeizer@wins.uva.nl

# 1. Introduction

The superconducting instability in heavy-electron compounds [1,2] continues to attract a great deal of attention. In the past years much research has been directed towards the close connection between superconductivity and magnetism in heavy-electron materials [3]. The principle research issues which have emerged are: (i) spin-fluctuation versus phonon mediated superconductivity, (ii) the symmetry of the superconducting gap function, and (iii) the interplay of magnetic order and superconductivity. Among the heavy-fermion superconductors $UPt_3$ with a superconducting transition temperature $T_c \sim 0.55$ K [4] is regarded as exemplary. The low-temperature normal state is characterized by pronounced antiferromagnetic spin-fluctuation phenomena ($T^* \sim 20$ K) and incipient magnetism [5], which give rise to a strong renormalization of the effective mass, of the order of 100 times the free electron mass. Neutron-diffraction experiments have shown that superconductivity in $UPt_3$ coexists with antiferromagnetic order, which develops below a Néel temperature $T_N \sim 6$ K [6]. The antiferromagnetic order is unconventional in the sense that the ordered moment squared $m^2(T)$ grows quasi-linearly with temperature. Moreover, the size of the ordered moment is extremely small $m = 0.02 \pm 0.01$ $\mu_B$/U-atom. The superconducting ground state is difficult to reconcile with strong magnetic interactions and, therefore, it has been suggested that superconductivity is mediated by antiferromagnetic interactions rather than by phonons [7]. However, decisive experimental evidence for this is still lacking. More recently, it has been argued that superconductivity is a more general property of heavy-fermion antiferromagnets close to a quantum critical point [8]. In the case of $UPt_3$ the quantum critical point might be reached by doping [5], but the concurrent non-Fermi-liquid behavior has not been signaled so far. In the past decade, evidence has accumulated that superconductivity in $UPt_3$ is truly unconventional, i.e. the symmetry of the superconducting gap function is lower than that of the underlying Fermi surface [9]. Evidence for this is in part presented by the power-law temperature dependence of the electronic excitation spectrum below $T_c$, indicating point nodes and/or line nodes in the gap [10]. The discovery of a multicomponent superconducting phase diagram with three vortex phases in the field-temperature plane [11-14], and the subsequent explanation within the Ginzburg-Landau theory of second order phase transitions (see Ref. 15 and references therein) is in general considered as hard proof for unconventional superconductivity.

UPt$_3$ is the only known superconductor with three different superconducting vortex phases. In zero magnetic field two superconducting phases are found, the A phase below $T_c^+ = 0.54$ K and the B phase below $T_c^- = 0.48$ K. In a magnetic field the A phase is suppressed, while the B phase transforms into a third phase, labeled C. The three phases meet in a tetracritical point. The phenomenology of the phase diagram has been studied extensively using Ginzburg-Landau (GL) theory, where the free energy functional is derived exclusively by symmetry arguments (the symmetry group for UPt$_3$ is $D_{6h}$). A number of GL models have been proposed [15-22] in order to explain the zero-field splitting $\Delta T_c = T_c^+ - T_c^-$ [11] and the topology of the phase diagram in magnetic field [12-14] or under pressure [23, 24]. Most of the GL models require an unconventional superconducting order parameter. The most plausible GL models which have been worked out to understand the phase diagram of UPt$_3$ fall into three categories: (i) the degeneracy of a two-dimensional (2D) even or odd parity order parameter is lifted by a symmetry breaking field [16-19], (ii) the spin degeneracy of a one-dimensional (1D) odd parity order parameter is lifted by a symmetry breaking field under the assumption of a weak spin-orbit coupling [15, 20], and (iii) there is an accidental degeneracy of two nearly degenerate 1D representations [19, 21]. However, no consensus has been reached, as each of the three models only partially describes the field and pressure variation of the superconducting phases. As regards the first two scenarios a key issue is to identify the symmetry breaking field (SBF). Experimental evidence that the anomalous weak antiferromagnetic order which sets in at $T_N \sim 6$ K acts as the SBF is at hand [25]. Another candidate for the SBF is the incommensurate structural modulation which has been detected by transmission electron microscopy [26]. However, its precise role remains unexplored.

In this paper we focus on the GL models with the degeneracy lifted by a SBF [15-20]. More specifically we investigate the role of the small-moment magnetism as SBF. Within the model (see section 2), the splitting of the superconducting transition temperature is proportional to the strength of the SBF or $\Delta T_c \propto \varepsilon$, where $\varepsilon \propto m^2$. Direct evidence for the coupling between $\Delta T_c$ and $m^2$ was deduced from specific-heat [23] and neutron-diffraction [25] experiments under hydrostatic pressure. It was observed that both $\Delta T_c$, determined by specific heat, and $m^2(T_c)$, measured by neutron diffraction under pressure, vary linearly with pressure and vanish at a critical pressure $p_c \sim 3$ kbar. We utilize another route to verify the coupling between $\Delta T_c$ and $m^2$, namely by doping UPt$_3$ with small amounts of Pd.

Vorenkamp and co-workers carried out specific-heat experiments on polycrystalline samples of $U(Pt_{1-x}Pd_x)_3$ ($x \leq 0.002$) and showed that $\Delta T_c$ almost doubles with respect to pure $UPt_3$ for the $x = 0.002$ compound [27]. This then directly prompted the question whether the enhancement of $\Delta T_c$ is due to the increase of the ordered moment $m$. Since it was known that for $0.02 < x < 0.07$ pronounced phase-transition anomalies in the thermal and transport properties signal an antiferromagnetic phase transition of the spin-density-wave type [5], we conducted a neutron-diffraction study on single-crystalline samples in order to investigate $m$ as function of Pd content over a wider range of $x$, including the region where $\Delta T_c$ is observed to increase. These results are reported in Ref. 28 and the conclusions are as follows. The small-moment antiferromagnetic order (SMAF) is robust upon doping with Pd and persists until at least $x = 0.005$. The ordered moment grows from $0.018 \pm 0.002$ $\mu_B$/U-atom for pure $UPt_3$ to $0.048 \pm 0.008$ $\mu_B$/U-atom for $x = 0.005$. For the SMAF $T_N \sim 6$ K and does not vary with Pd contents. For $x \geq 0.01$ a second antiferromagnetic phase is found, for which at optimum doping ($x = 0.05$) $T_N$ attains a maximum value of 5.8 K and the ordered moment equals $0.63 \pm 0.05$ $\mu_B$/U-atom. For this large-moment antiferromagnetic order (LMAF) $T_N(x)$ follows a Doniach-type diagram. From this diagram it is inferred that the antiferromagnetic instability in $U(Pt_{1-x}Pd_x)_3$ is located in the range 0.5-1.0 at.% Pd.

In this paper we present a study of the superconducting properties of $U(Pt_{1-x}Pd_x)_3$. The main objectives of this work are: (i) to determine $\Delta T_c(x)$ by means of specific-heat experiments, (ii) to test the SBF model by relating $\Delta T_c(x)$ to the ordered moment $m(x)$, and (iii) to investigate the effect of Pd doping on the superconducting phase diagram in the $B$-$T$ plane by means of magnetotransport and dilatometry experiments. The paper is organized as follows. In section 2 we review the basic relations for the SBF scenario. In section 3 we concentrate on the sample preparation process and the characterization of the samples by means of electrical resistivity. In section 4 we present and analyze the specific-heat of $U(Pt_{1-x}Pd_x)_3$ in the vicinity of the double superconducting transition. In sections 5 and 6 we present the magnetoresistance, thermal expansion and magnetostriction data and in section 7 we construct the phase diagrams in the $B$-$T$ plane for $x = 0.002$. In section 8, we extract the Ginzburg-Landau parameters, while the SBF model is tested in section 9. Finally, we present the concluding remarks in section 10. Parts of these results have been reported in a preliminary form in Refs. 29-31.

## 2. The SBF scenario

The SBF scenarios can be divided into two categories: (i) the degeneracy of a 2D even or odd parity order parameter is lifted by a SBF [16-19] and (ii) the spin degeneracy of a 1D odd parity order parameter is lifted by a SBF under the assumption of weak spin-orbit coupling [15, 20]. The irreducible representations for the superconducting gap with the appropriate $D_{6h}$ symmetry of UPt$_3$ have been tabulated by e.g. Yip and Garg (Ref. 32). We first concentrate on the 2D representation called the E-representation model. For a 2D representation with even parity, $E_{1g}$ or $E_{2g}$, or odd parity, $E_{1u}$ or $E_{2u}$, the superconducting gap function is given by $\Delta_E(\mathbf{k}) = \eta_x \Gamma_{E,x}(\mathbf{k}) + \eta_y \Gamma_{E,y}(\mathbf{k})$, where $\Gamma_{E,x}$ and $\Gamma_{E,y}$ are the basis functions for the relevant 2D-representation. The complex vector $\boldsymbol{\eta} = (\eta_x, \eta_y) = (|\eta_x|e^{i\varphi_x}, |\eta_y|e^{i\varphi_y})$ determines the order parameter. The free energy functional can be written as the sum of three terms [15-20]:

$$F = F_S + F_M + F_{SM} \tag{1}$$

Here $F_S$ is the free energy functional of the superconductor

$$F_S = \alpha_S (T - T_c)|\boldsymbol{\eta}|^2 + \frac{1}{2}\beta_1|\boldsymbol{\eta}|^4 + \frac{1}{2}\beta_2\left|\boldsymbol{\eta}^2\right|^2 \tag{2}$$

where the coefficients $\alpha_S$, $\beta_1$ and $\beta_2$ are stability parameters. The contribution from the magnetic order to the free energy is given by

$$F_M = \alpha_M (T - T_N) m^2 + \frac{1}{2}\beta_M m^4 \tag{3}$$

where $\mathbf{m} = (m,0,0)$ is the small ordered moment oriented along a principal axis in the basal plane ($T \le T_N$) and $\alpha_M$ and $\beta_M$ are stability parameters. The mixing term of magnetic order and superconductivity can be written as [16-19]:

$$F_{SM} = -\gamma m^2\left(\eta_x^2 - \eta_y^2\right) \tag{4}$$

where $\varepsilon = \gamma m^2$ is the symmetry breaking field. By minimizing the free energy it follows that the single superconducting transition at $T_c$ splits into two transitions at $T_c^+$ and $T_c^-$, where

$$\Delta T_c = T_c^+ - T_c^- = g\frac{|\gamma|}{\alpha_S}\frac{\beta_1 + \beta_2}{\beta_2}m^2 \tag{5}$$

Here $g = 1$ for the E-model. The ratio $(\beta_1+\beta_2)/\beta_2$ can be determined from the measured step sizes in the specific heat at $T_c^+$ and $T_c^-$:

$$\frac{\Delta c(T_c^-)/T_c^-}{\Delta c(T_c^+)/T_c^+} = 1 + \frac{\beta_2}{\beta_1} \tag{6}$$

Here the steps in the specific heat are measured relative to the normal state. The weak-coupling value for $\beta_2/\beta_1$ is 0.5.

The 1D representation model with odd parity yields very similar expressions. The three component order parameter is $\eta = (\eta_x, \eta_y, \eta_z) = (|\eta_x|e^{i\phi_x}, |\eta_y|e^{i\phi_y}, |\eta_z|e^{i\phi_z})$ and the gap function is given by [20]:

$$\Delta(\mathbf{k}) = \sum_{\lambda=x,y,z} \eta_\lambda l(\mathbf{k})\tau_\lambda \tag{7}$$

with $\tau_\lambda = i\sigma_y\sigma_\lambda$, where the $\sigma$'s denote the Pauli spin matrices. The complex coefficients $\eta_\lambda$ are characterized by a spin index $\lambda$. The orbital part, $l(\mathbf{k})$, belongs to the 1D representation $A_{2u}$, $B_{1u}$ or $B_{2u}$. The free energy functional is expressed as in the E-model using equations 1-3. The coupling term of the magnetic and the superconducting order parameter consists of three components and equation 4 now reads:

$$F_{SM} = -\gamma m^2 \left(2\eta_x^2 - \eta_y^2 - \eta_z^2\right) \tag{8}$$

For the 1D model $\Delta T_c$ is given by equation 5 with $g = 3/2$, while the expression for $\beta_2/\beta_1$ is the same as in the E-model (Eq. 6). Note that in Ref. 20 an incorrect expression is given for $T_c^-$ in which $\beta_1$ and $\beta_2$ are interchanged in the numerator.

As it is our purpose to verify Eq. 5 by experiments, one also needs, besides values for $\Delta T_c$ and $\beta_2/\beta_1$, which can be deduced from the specific-heat data, and the value of $m$, which follows from the neutron-diffraction experiments [28], an estimate for the parameter $|\gamma|/\alpha_S$. We comment on this in section 9.

## 3. Experimental

The data reported in this paper have been taken on annealed polycrystalline ($x \leq 0.006$) and single-crystalline samples ($x \leq 0.002$). Polycrystalline material was prepared by arc-melting the constituents in a stoichiometric ratio in an arc furnace on a water-cooled copper crucible under a continuously Ti-gettered argon atmosphere (0.5 bar). As starting materials we used uranium (JRC-EC, Geel) with a purity of 99.98%, and platinum and palladium (Johnson Matthey) with purity 5N. Polycrystalline material with low Pd contents ($x \leq 0.01$) was prepared by using master alloys (e.g. 5 at.% Pd). Single-crystals with $x = 0$ and 0.002 were prepared in a mirror furnace (NEC-NSC35) using the horizontal floating zone method. A single-crystalline sample with $x = 0.001$, was pulled from the melt using a modified Czochralski technique in a tri-arc furnace under a continuously Ti-gettered argon atmosphere. For annealing, the samples were wrapped in tantalum foil and put in a water-free quartz tube together with a piece of uranium that served as a getter. After evacuating ($p < 10^{-6}$ mbar) and sealing the tube, the samples were annealed at 950 °C during 4 days. Next the samples were slowly cooled in 3 days to room temperature. Several samples were investigated by Electron Probe Micro Analysis (EPMA), but the concentration of Pd is too small to arrive at a quantitative composition analysis. In the following, the value of $x$ is the nominal composition. Samples with appropriate dimensions and weights were obtained by means of spark erosion.

In order to characterize the prepared materials the electrical resistivity, $\rho(T)$, was measured on bar-shaped samples. The results for the polycrystalline samples ($x = 0, 0.0025, 0.003, 0.0035, 0.004$ and $0.005$) are reported in Ref. 30. The data above $T_c^+$ are well-described by the Fermi liquid expression $\rho = \rho_0 + AT^2$ ($T < 1$ K). The residual resistivity, $\rho_0$, is extracted by extrapolating the $AT^2$ term to $T = 0$. For $x = 0$, the residual resistance ratio RRR $= R(300\text{K})/R(0) \approx 1000$, indicating a high quality of the pure compound, while $T_c^+ = 0.57$ K. Upon alloying, $\rho_0$ increases linearly with $x$ at a rate of $9.6 \pm 0.2$ μΩcm/at.%Pd, which ensures that palladium dissolves homogeneously in the matrix. Also $T_c^+$ varies smoothly with Pd content and the critical concentration for the suppression of superconductivity is $x_{c,sc} \sim 0.007$. In case of the single-crystalline materials, $\rho(T)$ was obtained for a current, $I$, along the a and c-axis. The residual resistivity, $\rho_{0,a}$, amounts to 0.52, 1.6 and 2.5 μΩcm, while $\rho_{0,c}$, amounts to 0.18, 0.75 and 1.02 μΩcm, for $x = 0$, 0.001 and 0.002, respectively. For pure UPt$_3$ we obtain RRR values of ~460 and ~720 for $I \| \mathbf{a}$ and $I \| \mathbf{c}$, respectively. $T_c^+$ is suppressed at a rate 0.77 K/at.%Pd. In the following section we compare the resistively determined $T_c^+$ with the bulk value determined by the specific heat.

## 4. Specific heat of U(Pt$_{1-x}$Pd$_x$)$_3$

The specific heat, $c(T)$, of a series of U(Pt$_{1-x}$Pd$_x$)$_3$ samples was measured using the relaxation technique. Experiments have been carried out on annealed single-crystalline samples with $x = 0.000$, 0.001 and 0.002 and on annealed polycrystalline samples with $x = 0.000$, 0.0025, 0.003, and 0.004. The typical sample mass was 80 mg. The results are shown in Fig. 1 in a plot of $c/T$ versus $T$. At least three interesting features strike the eye: (i) $T_c^+$ and $T_c^-$ are well resolved for $x \leq 0.003$, while for $x = 0.004$ only $T_c^+$ is resolved (T> 0.1 K), (ii) both $T_c^+$ and $T_c^-$ decrease smoothly with Pd concentration, while $\Delta T_c$ increases, and (iii) the overall height of the jumps in $c/T$ at $T_c^+$ and $T_c^-$ decreases with increasing $x$. The results for the single-crystalline samples (x $\leq$ 0.002) are in good agreement with those obtained by Vorenkamp et al. [27] on polycrystalline material. In order to determine the ideal values for the jumps in the specific heat, we have made use of an equal entropy construction at the NA and AB phase boundaries. The ideal transitions are represented by the solid lines in Fig. 1. The resulting values of $T_c^+$, $T_c^-$, $\Delta T_c$, $\Delta_{NA}c(T_c^+)/T_c^+$, $\Delta_{NB}c(T_c^-)/T_c^-$, $\Delta_{AB}c(T_c^-)/T_c^-$ and $\beta_2/\beta_1$ are collected in Table I. Here the subscripts NA and NB refer to the step sizes measured with respect to the normal state $c/T$ value, while the subscript AB refers to the step size measured with respect to the $c/T$ value in the A phase. Below $T_c^-$ $c_s(T) = \gamma_0 T + \delta T^2$, down to the lowest $T$ measured (0.1 K). The $\delta T^2$ term shows that the superconducting gap function has a line node [9]. For $T \to 0$ K considerable residual $\gamma_0$ values are observed which is attributed to impurity broadening of the line node [33]. Just as is the case for pure UPt$_3$ [11,12], Fig. 1 shows that the superconducting state entropy exceeds the entropy of the normal state (assuming $c_N = \gamma_N T$). The extrapolated entropy unbalance for $0 < T < T_c^+$ is slightly sample dependent in U(Pt$_{1-x}$Pd$_x$)$_3$ and ranges from 6 to 12% of the normal state entropy. The entropy discrepancy can be resolved by either an increase of $c_N/T$ or a decrease of $c_s/T$ below 0.1 K. The most plausible explanation for the entropy imbalance is offered by the presence of an anomaly at 0.018 K in the normal state specific heat [34]. The entropy balance is fulfilled when this peak is included.

In Fig. 2 $T_c^+$, $T_c^-$ and $\Delta T_c$ are plotted as function of Pd concentration. Both, $T_c^+$ and $T_c^-$ decrease with increasing Pd concentration, but with different rates: d$T_c^+$/d$x$= -0.79±0.04 K/at.%Pd and d$T_c^-$/d$x$= -1.08±0.06 K/at.%Pd, and as a result $\Delta T_c$ increases at a rate d$\Delta T_c$/d$x$= 0.30±0.02 K/at.%Pd. The value d$T_c^+$/d$x$= -0.79±0.04 K/at.%Pd measured by the specific heat is within the experimental error equal to the resistive value -0.77 K/at.%Pd.

Usually, the ratio $\beta_2/\beta_1$ is calculated from the steps $\Delta c/T$ at $T_c^+$ and $T_c^-$ with respect to the normal phase (Eq. 6). However, in order to obtain proper values of $\beta_2/\beta_1$ one should realize that Eq. 6 is only correct for small values of $\Delta T_c$, which is not the case in the doped samples. Therefore we use a slightly different relation for $\beta_2/\beta_1$ given here below. The steps $\Delta c/T$ are derived from the GL free energy by $\Delta c/T = -\partial^2 F/\partial T^2$. The thermodynamic step in the specific heat at $T_c^-$ can be written as:

$$\Delta_{AB} c(T_c^-) / T_c^- = \Delta_{NB} c(T_c^-) / T_c^- - \Delta_{NA} c(T_c^+) / T_c^+ \qquad (9)$$

The temperature dependence of $c/T$ at the two phase transitions is shown schematically in Fig. 3. In the 4th order GL model $c/T$ is constrained to be temperature independent (solid line), while the measured behavior shows a clear temperature dependence (dotted line). In fact, higher order terms need to be taken into account in order to arrive at a temperature dependent $c/T$. In order to arrive at a proper estimate of $\beta_2/\beta_1$ we use the directly measured step $\Delta_{AB} c(T_c^-) / T_c^-$, instead of $\Delta c/T$ at $T_c^-$ with respect to the normal state. This results in:

$$\frac{\beta_2}{\beta_1} = \frac{\Delta_{AB} c(T_c^-) / T_c^-}{\Delta_{NA} c(T_c^+) / T_c^+} \qquad (10)$$

The values of $\beta_2/\beta_1$, determined from Eq. 10, are listed in Table I, and are close to the weak coupling limit 0.5. The ratio $\beta_2/\beta_1$ is within the experimental error independent of Pd concentration. Note that in a first analysis of the specific-heat data we used Eq. 6 which led to a steady decrease of $\beta_2/\beta_1$ upon Pd doping [29].

## 5. The upper critical field

In order to investigate the effect of Pd doping on the upper-critical field, $B_{c2}(T)$, we have measured the electrical resistivity in field for single-crystalline U(Pt$_{1-x}$Pd$_x$)$_3$ with $x = 0.001$ and 0.002. These experiments were primarily conducted to investigate the presence of a kink in $B_{c2}(T)$, which locates the tetracritical point in the multicomponent $B$-$T$ phase diagram of pure UPt$_3$. The experiments were carried out on bar-shaped samples with the current along the long axis ($\mathbf{I} \parallel \mathbf{a}$). The samples were cut from the same single-crystalline batch as used for the specific-heat (section 4) and neutron-diffraction experiments [28]. $B_{c2}(T)$ was determined by resistivity experiments in a transverse constant magnetic field for $\mathbf{B} \parallel \mathbf{c}$ and $\mathbf{B} \perp \mathbf{c}$ (i.e. $\mathbf{B} \parallel \mathbf{a}^*$, where $\mathbf{a}^*$ is taken at right angles to $\mathbf{a}$ and $\mathbf{c}$). In Fig. 4 some typical results are shown for $x = 0.001$ ($\mathbf{B} \perp \mathbf{c}$). In these low magnetic fields ($B < 1.3$ T) the magnetoresistance is small (less than 1% of $\rho_0$ per Tesla). At each applied field $T_c^+$ was determined by the 50% resistivity criterion, and the width of the superconducting transition, $\Delta T_c^+$, was determined by the 10-90% resistivity criterion. The resulting upper-critical field curves for $\mathbf{B} \perp \mathbf{c}$ and $\mathbf{B} \parallel \mathbf{c}$ are shown in Fig. 5, where both axes have been normalized by dividing by $T_c^+$. For comparison we have also plotted in Fig. 5 the resistively determined $B_{c2}(T)$ data of pure UPt$_3$ [35, 36].

For $\mathbf{B} \perp \mathbf{c}$ clear kinks in $B_{c2}(T)$ are observed (Fig. 5b). This strongly suggests that in the Pd doped samples ($x \leq 0.002$) a tetracritical point is present, as for pure UPt$_3$. Upon doping the tetracritical point shifts towards lower temperatures and higher fields, which indicates that the A phase becomes more stable. For $x = 0.001$ $T_{cr} = 0.309(8)$ K and $B_{cr} = 0.461(8)$ T, while for $x = 0.002$ $T_{cr} = 0.225(8)$ K and $B_{cr} = 0.490(8)$ ($\mathbf{B} \perp \mathbf{c}$). Thus for $\mathbf{B} \perp \mathbf{c}$ the phase diagrams for U(Pt$_{1-x}$Pd$_x$)$_3$ ($x \leq 0.002$) have the same topology. For $\mathbf{B} \parallel \mathbf{c}$ no distinct anomalies are observed in $B_{c2}(T)$ of the Pd doped samples (Fig. 5a). However, for pure UPt$_3$ a weak kink in $B_{c2}(T)$ was reported [36], locating the tetracritical point at $T_{cr} = 0.45(2)$ K and $B_{cr} = 0.60(2)$ T. In the following section we study the phase diagrams for $\mathbf{B} \perp \mathbf{c}$ and $\mathbf{B} \parallel \mathbf{c}$ in more detail by dilatometry.

## 6. Thermal expansion and magnetostriction

In order to determine the superconducting phase diagram of the $x = 0.002$ compound dilatometry experiments (thermal expansion and magnetostriction) have been performed. The results will be compared with the dilatation experiments on pure UPt$_3$ reported by Van Dijk et al. [37, 38].

### 6.1 Experimental

The U(Pt$_{0.998}$Pd$_{0.002}$)$_3$ sample used for the thermal expansion and magnetostriction experiments was cut from the same single-crystalline batch as prepared for the resistivity, specific heat and neutron-diffraction experiments. The approximate dimensions of the sample along the a, a$^*$ and c-axis are 4.0, 5.0 and 3.4 mm, respectively. The sample was mounted in a capacitance dilatation cell machined of oxygen-free high-conductivity copper. Two RuO$_2$ resistors which served as heater and thermometer were glued onto the sample. Length changes along the c-axis of the sample were determined with the three-terminal capacitor technique, using an Andeen-Hagerling bridge (model 2500E). The sensitivity of the experimental set-up amounts to 0.01 Å. The dilatation cell was attached to the cold finger of a dilution refrigerator.

The coefficient of linear thermal expansion, $\alpha = L^{-1}dL/dT$, was measured using a modulation technique ($f = 0.003$ Hz, $\Delta T = 5\text{-}10$ mK). The linear magnetostriction, $\lambda = (L(B)-L(0))/L(0)$, was measured by sweeping the magnetic field at a relatively low rate ($dB/dt \le 0.03$ T/min) while monitoring the length, $L$, of the sample. The magnetostriction was measured for a field along the dilatation direction ($\mathbf{B} \| \mathbf{c}$) and at right angles ($\mathbf{B} \| \mathbf{a}$).

## 6.2  Thermal expansion

The zero-field temperature variation of the coefficient of linear thermal expansion along the c-axis, $\alpha_c(T)$, of U(Pt$_{0.998}$Pd$_{0.002}$)$_3$ is shown in Fig. 6, with as inset $\alpha_c(T)/T$. Just as for pure UPt$_3$, $\alpha_c/T$ is constant in the normal state, while two clear steps of opposite sign (most pronounced in $\alpha_c(T)/T$) mark the double superconducting transition. The superconducting transition temperatures have been determined using an equal-length construction and the idealized transition is given by the solid line in Fig. 6. For $x = 0.002$, $T_c^+ = 0.381(2)$ K and $T_c^- = 0.271(4)$ K. These values are in excellent agreement with the transition temperatures $T_c^+ = 0.384(3)$ K and $T_c^- = 0.276(4)$ K determined by the specific heat (see section 4). However, the value of $T_c^+ = 0.420(3)$ K determined resistively is slightly higher. This has also been noticed for pure UPt$_3$ [12]. The resistive transition temperature marks the onset of the bulk transitions measured by the specific heat and thermal expansion. The difference between the resistive and bulk transition decreases in an applied magnetic field. In Fig. 7, a few exemplary $\alpha_c(T)$-curves are shown in a constant magnetic field ($\mathbf{B} \| \mathbf{c}$ and $\mathbf{B} \| \mathbf{a}$). Both $T_c^+$ and $T_c^-$ are suppressed with field, but $T_c^+$ is suppressed more rapidly than $T_c^-$, so that they merge at a critical field, $B_{cr}$. The field dependence is anisotropic. For $\mathbf{B} \| \mathbf{a}$ the transitions merge in the field range 0.5-0.6 T, while for $\mathbf{B} \| \mathbf{c}$ the transitions do merge at about 1.0 T, which is close to $B_{c2}$ at our lowest temperature (0.075 K).

Combining the thermal-expansion and the specific-heat data we can determine the uniaxial pressure dependence of the superconducting phase transitions using the Ehrenfest relation:

$$\frac{dT}{dp_i} = \frac{V_m \Delta\alpha_i}{\Delta(c/T)} \qquad (11)$$

Here $p_i$ is the uniaxial pressure along a specific crystallographic axis ($i = $ a, a$^*$, c) and $V_m = 4.24 \times 10^{-5}$ m$^3$/mol is the molar volume. With help of the thermal-expansion data of Fig. 6 and the specific-heat steps listed in Table I, we calculate: $dT_c^+/dp_c$ = -0.14(1) K/GPa and $dT_c^-/dp_c$ = 0.06(1) K/GPa. Thus for uniaxial pressure along the c-axis the splitting, $\Delta T_c = T_c^+ - T_c^-$, decreases at a rate $d\Delta T_c/dp_c$ = -0.20(2) K/GPa. These calculated pressure dependencies are similar to those determined from the pressure dependence of the specific heat for pure UPt$_3$, where $dT_c^+/dp_c$ = -0.13(3) K/GPa, $dT_c^-/dp_c$ = 0.09(3) K/GPa and $d\Delta T_c/dp_c$ = -0.22(6) K/GPa [39]. Assuming a linear pressure dependence of $T_c^+$ and $T_c^-$, the A phase vanishes at $p_{cr}$ = 0.54 GPa and $T_{cr}$ = 0.308 K for $x$ = 0.002, while for pure UPt$_3$ $p_{cr}$ is only 0.25 GPa, because of the much smaller zero-pressure splitting, and $T_{cr}$ = 0.459 K.

## 6.3 Magnetostriction

The linear magnetostriction along the c-axis, $\lambda_c(B)$, at $T$ = 0.075 K is shown in Fig. 8 for fields up to 2 T (**B**∥ **c** and **B**∥ **a**). In addition to the normal state contribution to $\lambda_c(B)$, a superconducting contribution is present below $B_{c2}$. For **B**∥ **c** the normal-state magnetostriction (B< 2 T) is well described by a quadratic field dependence, $\lambda_c(B) = \lambda_c(0) + b_c B^2$. The coefficient of the quadratic term, $b_c$, is slightly temperature dependent and is -1.66x10$^{-7}$ T$^{-2}$ at $T$ = 0.075 K. In Fig. 9 we show $\lambda_c(B)$ with **B**∥ **a** at several temperatures. Here the normal-state magnetostriction also follows a $B^2$ dependence (B< 2 T) with $b_c$ is -0.46x10$^{-7}$ T$^{-2}$ at $T$ = 0.075 K. For **B**∥ **a** the upper critical field $B_{c2}$ is difficult to distinguish, while the B-to-C phase transition at $B_c^*$ is visible as a clear kink in the data. For **B**∥ **c** the situation is reversed: $B_{c2}$ shows up as a clear anomaly on the $\lambda_c(B)$ curve, while $B_c^*$ does not. For this field direction, the superconducting signal, obtained after subtracting the quadratic background contribution, is show in Fig. 10. Although the behavior observed for $x$ = 0.002 is in many aspects similar to the behavior for pure UPt$_3$, two important differences should be noted: (i) for $x$ = 0.002 only $B_{c2}$ is resolved from the magnetostriction curves for **B**∥ **c**, while for pure UPt$_3$ both $B_{c2}$ and $B_c^*$ are resolved, and (ii) for $x$ = 0.002 a significant hysteresis is observed for **B**∥ **c**, which was absent in the data of pure UPt$_3$. In Fig. 11 a typical magnetostriction cycle (field sweep up and down) is shown. In Fig. 12 we show for both field orientations the amount of hysteresis, obtained after subtracting the sweep-up signal from the sweep-down signal. For **B**∥ **a** the hysteresis is negligible, while for **B**∥ **c** the resulting curve has two peaks. The peak just below $B_{c2}$ is reminiscent of the peak effect observed in metallic alloys with strong pinning of the flux-line lattice. Recently, the peak effect was found in the magnetization of several pure UPt$_3$ samples [40,41]. The peak effect is expected to become more pronounced upon introducing additional pinning centers, e.g. by doping with Pd. The larger low-field peak, observed for $x=0.002$ at $B\sim 0.1$ T, which is most pronounced for $\mathbf{B}\parallel \mathbf{c}$ (see Fig. 12a), has also been reported for pure UPt$_3$. This peak, which has a weak temperature dependence, is not directly related to the superconducting properties as it is also present in the normal state. The origin of this anomaly remains unclear, but it has been suggested that it is related to a meta-stable magnetic state [42].

With help of the measured discontinuities at the superconducting transitions the field dependence of the transition temperatures can be estimated with the Ehrenfest relation:

$$\left(\frac{\partial T_c}{\partial B}\right)_p = -\frac{\Delta\tau_i}{\Delta\alpha_i} \tag{12}$$

Here $\tau_i = \mathrm{d}\lambda_i/\mathrm{d}B$ where $i$ refers to the principal crystallographic directions. The initial field dependencies of $T_c^+$ and $T_c^-$ are determined by the $B=0$ thermal expansion data and by extrapolation of $\Delta\tau_c$ to $B\rightarrow 0$. The values determined in this way are listed in Table II and should be compared to the directly measured slopes of the phase lines of the superconducting phase diagram of U(Pt$_{0.998}$Pd$_{0.002}$)$_3$ (see section 8).

# 7. Superconducting phase diagram of U(Pt$_{0.998}$Pd$_{0.002}$)$_3$

By locating the anomalies at the superconducting phase transitions determined by our dilatometry experiments in the $B$-$T$ plane we have constructed the superconducting phase diagrams of U(Pt$_{0.998}$Pd$_{0.002}$)$_3$ shown in Fig. 13. The NA transition is detected by both thermal expansion and magnetostriction, while the AB transition shows up only in the thermal expansion. The NC phase line, which is only found for $\mathbf{B}\perp \mathbf{c}$, has a very weak signature in the thermal-expansion data and was therefore complemented by the $B_{c2}(T)$ data measured resistively (Fig. 5b). In this field range the resistive and bulk $T_c^+$ are equal within the experimental accuracy. For $\mathbf{B}\perp \mathbf{c}$ the tetracritical point is located at $T_{cr}= 0.205(4)$ K and $B_{cr}= 0.556(8)$ T. For $\mathbf{B}\parallel \mathbf{c}$ the $x=0.002$ compound has no tetracritical point ($T\geq 0.075$ K), which presents a striking difference with respect to pure UPt$_3$.

The AB phase line of U(Pt$_{0.998}$Pd$_{0.002}$)$_3$ measured for $\mathbf{B}\perp \mathbf{c}$ shows a clear change of slope at $B=0.2$ T. For pure UPt$_3$ a similar kink was observed, albeit at a lower field $B=0.1$ T. It has been suggested that this kink arises from a coupling of the superconducting order parameter to the meta-stable magnetic state [42]. For U(Pt$_{0.998}$Pd$_{0.002}$)$_3$, however, the change of slope does not coincide with the low-field anomaly observed in the magnetostriction at $B$= 0.1 T, and the origin remains unclear.

In Table III we compare the measured slopes of the phase lines near the tetracritical point with the calculated ones using the Ehrenfest relation (Eq.12). Within the experimental accuracy the data agree, which demonstrates their internal consistency. Near the tetracritical point the thermodynamic steps should follow the relation $\Delta_{NA} + \Delta_{AB} = \Delta_{NC} + \Delta_{BC}$, where $\Delta$ is $\Delta c/T$, $\Delta\alpha$ or $\Delta\tau$. We have checked that this relation holds for $\Delta\tau$. The thermodynamic stability of a phase diagram with a tetracritical point, where at least three second order phase-transition lines meet, leads to strict conditions for the slopes of the four phase lines as formulated in Ref. 43. In the case of pure UPt$_3$ these conditions were satisfied [38]. In order to investigate the thermodynamic stability of U(Pt$_{0.998}$Pd$_{0.002}$)$_3$ additional specific-heat measurements in an applied magnetic field are needed.

## 8.  Ginzburg-Landau parameters of U(Pt$_{1-x}$Pd$_x$)$_3$

The temperature derivative of the thermodynamic critical field, d$B_c$/d$T$, near $T_c$, can be estimated from the jump in the specific heat at the superconducting transition in zero field [44]:

$$\Delta(c/T) = \frac{V_m}{\mu_0}\left(\frac{\partial B_c}{\partial T}\right)^2 \qquad \textbf{(13)}$$

Here $\mu_0$ the permeability of free space. The thermodynamic critical field is related to the upper critical field by $B_{c2}$= $2\kappa B_c$, where $\kappa$ is the (isotropic) Ginzburg-Landau parameter which characterizes the superconducting state [44]. The GL parameter is defined as $\kappa = \lambda/\xi$, where $\lambda$ is the penetration depth and $\xi$ is the coherence length. Since we are dealing with a hexagonal strongly anisotropic material $\kappa$, $\lambda$ and $\xi$ are anisotropic. In that case, the upper critical field is given by $B_{c2}^i = \Phi_0 / (2\pi\xi_j\xi_k)$, where $\Phi_0$ is the flux quantum and $i$, $j$ and $k$ are the principal crystallographic directions. Assuming that the parameters in the basal plane are isotropic the following relations between $B_{c2}$ and $B_c$ hold:

$$\begin{aligned}B_{c2}^a &= \sqrt{2\kappa_a\kappa_c}\, B_c \\ B_{c2}^c &= \sqrt{2}\kappa_a B_c\end{aligned} \qquad \textbf{(14)}$$

Here the superscripts a and c refer to the direction of the magnetic field, $\kappa_a = \lambda_a/\xi_a$ and $\kappa_c = \lambda_c/\xi_c$. The average GL parameter is defined as $\kappa_{av} = (\kappa_a^2 \kappa_c)^{1/3}$.

We have evaluated the various GL parameters and the temperature derivatives of the upper critical fields from the measured data. The results for $B \rightarrow 0$ are listed in Table IV. The value of $dB_c/dT$ for the A phase has been determined using $\Delta_{NA}c(T_c^+)/T_c^+$. The values of $dB_{c2}/dT$ determined from the step in the magnetostriction and thermal expansion (see Table II) are in reasonable agreement with the values determined directly from the slope of the phase diagrams.

UPt$_3$ is an extreme type-II superconductor with for pure samples $\lambda \geq 6000$ Å and $\xi \sim 120$ Å so that $\kappa \sim 50$. Upon Pd doping ($x \leq 0.002$) $\kappa_{av}$ remains roughly constant, while $\kappa_a$ decreases and $\kappa_c$ increases (see Table IV). Substituting Pd makes the superconducting properties less anisotropic, which is also reflected in the ratio of the anisotropic quasiparticle masses, determined by $m_c/m_a = (B_{c2}^c/B_{c2}^a)^2$ (see Table IV). The lower-critical field, $B_{c1}$, is related to the thermodynamic critical field according to $B_{c1} = B_c \ln(\kappa)/(\sqrt{2}\kappa)$, from which it follows that $B_{c1}$ is about 6% of $B_c$. Such small values of $B_{c1}$ have not been probed in our dilatometry experiments.

## 9. Testing the SBF model

One of the main objectives of the specific-heat experiments on the U(Pt$_{1-x}$Pd$_x$)$_3$ system was to determine the superconducting splitting $\Delta T_c$ as function of $x$. From the data in Fig. 2 we conclude that $\Delta T_c$ increases linearly with $x$ at a rate $d\Delta T_c/dx = 0.30 \pm 0.02$ K/at.%Pd. Within the GL models presented in section 2, $\Delta T_c$ is proportional to the strength of the SBF or assuming that the ordered moment of the SMAF is the SBF $\Delta T_c \propto m^2(T_c^+)$ (see Eq. 5; we comment on the prefactor $(|\gamma|/\alpha_s)(\beta_1+\beta_2)/\beta_2$ at the end of this section). In order to determine $m^2(x)$ we have recently carried out neutron-diffraction experiments [28] on single-crystalline samples. The ordered moments at $T_c^+$ are 0.018(2), 0.024(3), 0.036(3) and 0.048(8) $\mu_B$/U-atom, for $x = 0$, 0.001, 0.002 and 0.005, respectively. In Fig. 14 we have traced $\Delta T_c$ as a function of $m^2(T_c^+)$ for $x \leq 0.002$. It is interesting to compare this result with $\Delta T_c \propto m^2(T_c^+)$ obtained by the hydrostatic pressure experiments [25], because doping increases $\Delta T_c$ and hydrostatic pressure decreases $\Delta T_c$. A direct comparison is not possible because of the relatively large uncertainty in the absolute value of $m^2(p=0)= 0.03\pm0.01$ $\mu_B$/U-atom. Therefore, we scaled the $m^2(p)$ values of Ref. 25 such that $m^2(p=0)= 0.02$ $\mu_B$/U-atom. The error bars for the pressure data correspond to the relative errors determined by counting statistics, while for the Pd doping these are absolute errors. After including the pressure data in Fig. 14, we notice the following three points: (i) both the Pd doping and pressure data sets collapse onto one curve, (ii) $\Delta T_c$ is a smooth function of $m^2(T_c^+)$, and (iii) $\Delta T_c \propto m^2(T_c^+)$ but in a limited range $\Delta T_c \leq 0.05$ K. The latter result shows that the simple GL models presented in section 2 break down for splittings $\Delta T_c > 0.05$ K. This is not unrealistic because the applied Ginzburg-Landau expansion is only valid for $\Delta T_c/T_c \ll 1$. Clearly, for enhanced splittings a more sophisticated Ginzburg-Landau expansion with terms beyond 4th order is needed. We conclude that there is a clear correlation between $\Delta T_c$ and $m^2(T_c^+)$, which is in line with the SMAF acting as SBF.

## 10. Concluding remarks

The experimental results reported in sections 3-6 show that the unconventional superconducting properties of UPt$_3$ are extremely sensitive to Pd doping. First of all, resistivity experiments show that the A phase signaled by $T_c^+$ is completely suppressed at a critical concentration $x_{c,sc}$ is ~0.007. Secondly, the specific-heat experiments show that the B phase, marked by the second transition at $T_c^-$, is suppressed even more rapidly, with $x_c \sim 0.004$. Thirdly, $\Delta T_c$ increases with Pd contents. One of the main objectives of the present work was to investigate the GL model formulated in section 2, especially with respect to whether the SMAF acts as the SBF. Indeed, we find a close correlation between $\Delta T_c(x)$ and $m^2(T_c^+)$. However, only for $\Delta T_c \leq 0.050$ K the proportionality between $\Delta T_c$ and $m^2$, predicted by the SBF model, holds. For $\Delta T_c > 0.050$ K $m^2$ grows more rapidly. The failure of the model for larger splittings is attributed to the limited applicability of the simple GL E-representation and 1D odd parity models. The 4th order expansion near $T_c$ is only valid for $\Delta T_c/T_c \ll 1$.

While SMAF and superconductivity coexist, evidence is accumulating that LMAF and superconductivity compete. Recent neutron-diffraction [28] and μSR [45] experiments indicate that the critical concentration for the onset of LMAF is near $x_{c,af} \cong x_{c,sc} \sim 0.007$ [30]. In order to put this on firm footing additional μSR experiments are in progress. The competition between superconductivity and LMAF lends further support for spin-fluctuation mediated superconductivity.

The effects of Pd doping (this work) and hydrostatic pressure [23] on the stability of the A phase are opposite. It is interesting to note that this also holds for the B and C phases [24, 37, 38], which is most clearly observed for $\mathbf{B} \perp \mathbf{c}$. By applying hydrostatic pressure the tetracritical point in the $B$-$T$ plane shifts to lower fields. Upon increasing pressure first the A phase disappears (at $p_c \sim 0.35$ GPa), followed by the B phase, so that the C phase is the most stable phase under pressure [24, 37, 38]. For Pd doping the contrary takes place. Upon doping the tetracritical point shifts to higher fields, and the A phase gains stability at the expense of the B and C phases. Note that the C phase is completely suppressed for $U(Pt_{0.998}Pd_{0.002})_3$ in the case $\mathbf{B} \parallel \mathbf{c}$. The normal-state properties of $UPt_3$ react upon Pd doping also in an opposite way to hydrostatic pressure. Experiments demonstrate that doping of 1 at.% Pd corresponds to an external pressure of about -0.33 GPa [46, 47]. This illustrates that the change of the normal-state properties is not governed by the volume, because both Pd doping and applying pressure reduce the unit cell volume. Instead, these changes can be explained, to a certain extent, by the change in the c/a ratio. In the case of Pd substitution $\Delta(c/a)/(c/a) = -0.6 \times 10^{-4}$ per at.% Pd, while for hydrostatic pressure, because of the anisotropic compressibility ($\kappa_c < \kappa_a$), $\Delta(c/a)/(c/a) = 1.3 \times 10^{-4}$ per GPa [5], hence, doping 1 at.%Pd corresponds to an applied hydrostatic pressure of -0.5(2) GPa. For the stability range of the A phase we do not arrive at the same numbers. For pure $UPt_3$, $d\Delta T_c/dp = -0.19$ K/GPa [23], while $d\Delta T_c/dx = 0.30$ K/at.%Pd. Thus in this case 1 at.%Pd corresponds to -1.6 GPa.

In analyzing the specific-heat data around the double superconducting transition, we have provided evidence that magnetism provides the SBF. This restricts the choice of the GL models to the E-representation model, which applies for both even and odd parity states, and to the 1D odd parity model. The latter model relies on a weak spin-orbit coupling. In zero magnetic field both models give identical results, but they differ in predicting the field and pressure dependence of the superconducting phases. Notably, a tetracritical point for all field directions is only possible in the E-model under certain conditions and certain symmetries ($E_{1g}$ [48] or $E_{2u}$ [15]), while no additional constraints are needed in the odd-parity 1D model. As regards, the pressure dependence, the E model predicts the B phase to be the stable phase under pressure. A recent refinement of the odd-parity 1D model shows that the C phase is most stable under pressure [49]. This is in line with recent pressure studies [24] and dilatometry experiments [38]. Moreover, NMR experiments [50, 51] have demonstrated convincingly that (i) the Knight shift does not change through the normal-superconducting phase transition, and (ii) the effective spin orbit coupling is weak. All these studies provide a strong case for the odd parity 1D GL model. It is interesting to note that in the refined 1D odd parity model [49] the antiferromagnetic moment is not static but fluctuates in time. This is consistent with recent NMR [50], μSR [45,52] and neutron diffraction [28] experiments.

In summary we have studied the superconducting phase diagram of U(Pt$_{1-x}$Pd$_x$)$_3$, by (magneto)resistance, specific heat and dilatometry. Our results in zero field show a strong increase of the splitting $\Delta T_c$ as function of Pd concentration. $\Delta T_c(x)$ correlates with an increase of the magnetic moment $m(x)$ upon Pd doping. This provides further evidence for the Ginzburg-Landau scenario with magnetism as the symmetry breaking field. The tetracritical point in the *B-T* plane is robust upon alloying for **B**⊥ **c**, at least till $x = 0.002$, while it is rapidly suppressed for **B**∥ **c**. In a magnetic field the A and B phases gain stability at the expense of the C phase upon alloying. In this sense Pd doping and the effect of an external pressure are complementary.

## Acknowledgments

This work was part of the research program of the Dutch Foundation for Fundamental Research of Matter ("Stichting" FOM). Work at Boston College was supported through Research Corporation Grant RA0246. MJG and AdV acknowledge support through NATO Collaborative Research Grant CGR11096.

Table I    Parameters deduced from the specific heat of U(Pt$_{1-x}$Pd$_x$)$_3$. The ratio $\beta_2/\beta_1$ is calculated with help of Eq. 10. The superscripts s and p refer to single- and polycrystalline samples respectively.

| $x$ (%) | $T_c^+$ (K) | $T_c^-$ (K) | $\Delta T_c$ (K) | $\Delta_{NA}c(T_c^+)/T_c^+$ (J/mol K$^2$) | $\Delta_{NB}c(T_c^-)/T_c^-$ (J/mol K$^2$) | $\Delta_{AB}c(T_c^-)/T_c^-$ (J/mol K$^2$) | $\beta_2/\beta_1$ |
|---|---|---|---|---|---|---|---|
| 0.00$^p$ | 0.560(3) | 0.506(3) | 0.054(4) | 0.23(1) | 0.34(1) | 0.14(1) | 0.60(6) |
| 0.00$^s$ | 0.543(3) | 0.489(3) | 0.054(4) | 0.26(1) | 0.35(1) | 0.13(1) | 0.50(5) |
| 0.10$^s$ | 0.437(3) | 0.355(3) | 0.082(4) | 0.21(1) | 0.26(1) | 0.12(1) | 0.57(7) |
| 0.20$^s$ | 0.384(3) | 0.276(4) | 0.108(5) | 0.17(1) | 0.19(1) | 0.10(1) | 0.58(8) |
| 0.25$^p$ | 0.362(3) | 0.236(4) | 0.126(5) | 0.18(1) | 0.18(1) | 0.08(1) | 0.44(9) |
| 0.30$^p$ | 0.313(4) | 0.163(5) | 0.150(6) | 0.13(1) | 0.09(1) | 0.06(1) | 0.46(11) |
| 0.40$^p$ | 0.222(5) | - | - | 0.07(1) | - | - | - |

Table II    Thermodynamic quantities $\Delta\alpha_c$ and $\Delta\tau_c$ for $x = 0.002$ at the superconducting transitions $T_c^+$ and $T_c^-$ in zero field. The step $\Delta\tau_c$ is determined in the limit $B \to 0$.

| | $T_c^+$ (K) | $T_c^-$ (K) |
|---|---|---|
| $\Delta\alpha_c$ ($10^{-6}$K$^{-1}$) | -0.56(2) (NA) | 0.14(1) (AB) |
| $\Delta\tau_c$ ($10^{-6}$T$^{-1}$) | -0.11(1) ($\mathbf{B}\|\mathbf{a}$) <br> -0.07(1) ($\mathbf{B}\|\mathbf{c}$) | 0.01(1) ($\mathbf{B}\|\mathbf{a}$) <br> 0.01(1) ($\mathbf{B}\|\mathbf{c}$) |
| $-\Delta\tau_c/\Delta\alpha_c$ (K/T) | -0.20(2) ($\mathbf{B}\|\mathbf{a}$) <br> -0.13(1) ($\mathbf{B}\|\mathbf{c}$) | 0.07(7) ($\mathbf{B}\|\mathbf{a}$) <br> 0.07(7) ($\mathbf{B}\|\mathbf{c}$) |

Table III  Thermodynamic quantities for $x = 0.002$ in the vicinity of the tetracritical point ($\mathbf{B} \parallel \mathbf{a}$) at $T_{cr} = 0.205(4)$ K and $B_{cr} = 0.556(8)$ T. The phase lines between the A, B, C and N phases are indicated by NA, NC, AB and BC.

| | NA | NC | AB | BC |
|---|---|---|---|---|
| $dT/dB$ (K/T) | -0.472(8) | -0.217(4) | -0.086(1) | 1.37(6) |
| $\Delta\alpha_c$ ($10^{-6}$K$^{-1}$) | -0.18(2) | -0.02(2) | 0.06(1) | -0.10[*] |
| $\Delta\tau_c$ ($10^{-6}$T$^{-1}$) | -0.11(1) | 0.00(1) | 0.00(1) | -0.11(1) |
| $-\Delta\tau_c/\Delta\alpha_c$ (K/T) | -0.61(12) | 0.0(5) | 0.0(2) | 1.1(1) |

[*] Determined by the relation $\Delta_{NA} + \Delta_{AB} = \Delta_{NC} + \Delta_{BC}$.

Table IV  Slopes of the critical fields for the A phase of U(Pt$_{1-x}$Pd$_x$)$_3$, and the calculated GL parameters and effective mass ratio.

| $x$ (%) | $dB_c/dT$ (T/K) | $dT/dB_{c2}^a$ (K/T) | $dT/dB_{c2}^c$ (K/T) | $\kappa_a$ | $\kappa_c$ | $\kappa_{av}$ | $m_c/m_a$ |
|---|---|---|---|---|---|---|---|
| 0.0 | -0.087(6) | -0.241(8)[*] | -0.093(4)[*] | 87(6) | 13(1) | 46(3) | 6.7(6) |
| 0.1 | -0.078(6) | -0.250[**] | -0.124[**] | 73(6) | 18(1) | 46(3) | 4.1 |
| 0.2 | -0.071(7) | -0.258(6) | -0.155(6) | 64(6) | 23(2) | 46(3) | 2.7(2) |

[*] Data taken from ref. 38.
[**] Average value of the entries for the $x = 0.000$ and $x = 0.002$ compounds.

## Figure caption

Fig. 1    Specific heat divided by $T$ versus $T$ of U(Pt$_{1-x}$Pd$_x$)$_3$ for $x$= 0.000, 0.001 and 0.002 (single crystals) and for $x$= 0.0025, 0.003, and 0.004 (polycrystalline samples). The solid lines represent ideal transitions determined from an equal entropy construction.

Fig. 2    $T_c^+$, $T_c^-$ and $\Delta T_c$ of U(Pt$_{1-x}$Pd$_x$)$_3$ as a function of Pd concentration, determined from the specific-heat data. The solid and open symbols represent single- and poly-crystalline data, respectively.

Fig. 3    Schematic temperature dependence of $c/T$ at the double superconducting transition. The solid line represents $c/T$ calculated from the GL free energy, while the dotted line reflects the observed behavior.

Fig. 4    Resistivity of U(Pt$_{0.999}$Pd$_{0.001}$)$_3$ ($\mathbf{I} \| \mathbf{a}$) in constant magnetic fields $\mathbf{B} \| \mathbf{a}^*$, ranging from 0 to 1.3 T in steps of 0.1 T. For the most right curve $B$= 0 T and for the most left curve $B$= 1.3 T.

Fig. 5    The upper critical field of U(Pt$_{1-x}$Pd$_x$)$_3$, determined by resistivity ($\mathbf{I} \| \mathbf{a}$), in a plot of $B_{c2}/T_c$ as a function of $T/T_c$ for (a) $\mathbf{B} \| \mathbf{c}$ and (b) $\mathbf{B} \| \mathbf{a}^*$. $T_c = T_c^+$ is 0.547(5) K, 0.466(5) K and 0.420(5) K for $x$= 0.000, 0.001 and 0.002, respectively. The arrows mark the tetracritical points. The data of pure UPt$_3$ are taken from Refs. 35 and 36.

Fig. 6    The coefficient of linear thermal expansion along the c-axis ($\alpha_c$) of U(Pt$_{0.998}$Pd$_{0.002}$)$_3$. The solid line represents the ideal transition determined from an equal-length construction. In the inset the data is plotted as $\alpha_c/T$ vs. $T$. The transition temperatures are $T_c^+$= 0.381(2) and $T_c^-$= 0.271(4).

Fig. 7    The coefficient of linear thermal expansion along the c-axis ($\alpha_c$) of U(Pt$_{0.998}$Pd$_{0.002}$)$_3$ in magnetic fields ranging from 0 to 1 T as indicated, with (a) $\mathbf{B} \| \mathbf{c}$ and (b) $\mathbf{B} \| \mathbf{a}$. The curves in field are shifted upwards along the vertical axis for the sake of clarity.

Fig. 8  Linear magnetostriction along the c-axis ($\lambda_c$) of U(Pt$_{0.998}$Pd$_{0.002}$)$_3$ at $T$= 0.075 K for **B**$\parallel$ **a** and **B**$\parallel$ **c**. The dotted line for **B**$\parallel$ **c** represents the extrapolated normal state magnetostriction (see text).

Fig. 9  Linear magnetostriction along the c-axis ($\lambda_c$) with increasing magnetic field of U(Pt$_{0.998}$Pd$_{0.002}$)$_3$ for **B**$\parallel$ **a**. The temperature ranges from 0.10 to 0.35 K as indicated. The arrows mark the BC and CN transitions (see text). The curves for $T \geq 0.15$ K are shifted upwards for the sake of clarity.

Fig. 10  Linear magnetostriction along the c-axis ($\lambda_c$) of U(Pt$_{0.998}$Pd$_{0.002}$)$_3$ for increasing **B**$\parallel$ **c** at temperatures as indicated. The normal state contribution is subtracted (see text).

Fig. 11  Linear magnetostriction along the c-axis ($\lambda_c$) at $T$= 0.075 K of U(Pt$_{0.998}$Pd$_{0.002}$)$_3$ for **B**$\parallel$ **c**. The normal state contribution is subtracted. The arrows indicate the sweeps up and down of the magnetic field. The inset shows a close-up of the irreversible magnetostriction peak just below $B_{c2}$.

Fig. 12  The hysteresis in $\lambda_c$ for field sweeps up and down ($\lambda_{c,up}$-$\lambda_{c,down}$) of U(Pt$_{0.998}$Pd$_{0.002}$)$_3$ with (a) **B**$\parallel$ **a** and (b) **B**$\parallel$ **c**. The peak effect is observed just below $B_{c2}$ for **B**$\parallel$ **c**. The temperatures range between 0.075 and 0.4 K as indicated.

Fig. 13  The superconducting phase diagram of U(Pt$_{0.998}$Pd$_{0.002}$)$_3$ for **B**$\perp$ **c** and **B**$\parallel$ c, constructed from phase transitions detected in the thermal expansion (O) and magnetostriction (●). For **B**$\perp$ **c** the CN phase transition is determined resistively ($\Delta$).

Fig. 14  The variation of the splitting $\Delta T_c$ as a function of $m^2(T_c^+)$ for U(Pt$_{1-x}$Pd$_x$)$_3$ (●) and for UPt$_3$ under pressure ($\square$) [25]. For $\Delta T_c <$ 0.05 K $\Delta T_c \propto m^2$ as predicted by the GL model (see text). The solid line is to guide the eye.